\documentclass[a4paper,conference]{IEEEtran}

\usepackage{amsthm,amsmath,amssymb}

\ifCLASSINFOpdf
\else
\fi

\usepackage{url}

\newtheorem{mytheorem}{Theorem}

\newtheorem{mylemma}[mytheorem]{Lemma}
\newcommand{\fourier}[1]{\hat{#1}}
\newcommand{\Fourier}{\mathcal{F}}
\newcommand{\spec}{\text{spec}}
\newcommand{\trace}{\text{tr}\,}
\newcommand{\schattenclass}{\mathcal{I}}
\newcommand{\opnorm}{\text{op}}
\newcommand{\Order}{\mathcal{O}}
\newcommand{\Reals}{\mathbb{R}}
\newcommand{\Naturals}{\mathbb{N}}
\newcommand{\traceP}{\text{tr}_\alpha\,}

\renewcommand{\url}[1]{}

\begin{document}
%
\title{On the Szeg\"o--Asymptotics for Doubly--Dispersive Gaussian Channels}

\author{\IEEEauthorblockN{Peter Jung}
  \IEEEauthorblockA{
    TU Berlin,
    Einsteinufer 25, 10587 Berlin, Germany \\
    peter.jung@mk.tu-berlin.de}}


%


\maketitle

\begin{abstract}
We consider the time--continuous doubly--dispersive channel with additive Gaussian noise
and establish
a capacity formula for the case where the channel correlation operator is represented by a 
symbol which is periodic in time and fulfills some further integrability and smoothness conditions.
The key to this result is a new Szeg\"o formula for certain pseudo--differential operators.
The formula justifies the water--filling principle along time and frequency in terms
of the time--continuous time--varying transfer function (the symbol).
\end{abstract}


%
\IEEEpeerreviewmaketitle

\section{Introduction}
The information--theoretic treatment of the time--continuous  channel 
dispersive in time and frequency (doubly--dispersive) with additive
Gaussian noise has been a problem of long interest. 
A well known result for the time--invariant
and power--limited case has been achieved 
by Gallager and Holsinger \cite{Holsinger1964} and \cite{gallager:inftheo} in 
discretizing  the time--continuous problem into an increasing sequence of parallel memoryless channels
with known information capacity $I_n$. Coding theorems for the time--discrete Gaussian channel 
can be used for the time--continuous channel whenever such a discretization is realizable.
A direct coding theorem without discretization has been established by Kadota and Wyner \cite{Kadota1972}
for the causal, stationary and asymptotically memoryless channel. 

The discretization in \cite{gallager:inftheo}
was achieved by representing a single use of the time--continuous channel as the restriction of the channel
operator to time intervals $\alpha\Omega$ of length $\alpha$. The quantity $I_n$ is then determined by 
spectral properties of the restricted operator.
A major step in the calculation for the time--invariant case was the exact determination of the limit:
\begin{equation}
   I(S ):=\lim_{\alpha\rightarrow\infty}
   \left(\frac{1}{\alpha}\lim_{n\rightarrow\infty}I_n(\alpha S)\right)
   \label{eq:capacity:contin}
\end{equation}
which relies on the Kac--Murdock--Szeg\"o result \cite{Kac53} on the asymptotic spectral behavior of
convolution operators. As the classical result of Shannon for the time--continuous band--limited 
channel and the discussion in \cite{wyner:bandlimited:capacity} shows, $I(S)$ has only 
a meaning of coding capacity for given power budget $S$ whenever there exists a sequence
of nested intervals of length $\alpha_k$ (i.e. realizable discretization) approaching this limit as $k\rightarrow\infty$. Some remaining problems in this direction, like for example the 
robustness of this limit against interference between 
different blocks, have been resolved for Gallager--Holsinger model
in \cite{cordaro:intersymbol:gaussian}. The limit has the advantage
of nice interpretation as ''water--filling'' along the frequencies:
\begin{equation}
   I(S)=\int_{B\cdot\sigma(\omega)\geq 1}\log(B\cdot\sigma(\omega)) d\omega
   \label{eq:capacity:holsinger}
\end{equation}
where $\sigma$ denotes the symbol of the correlation operator $L_\sigma$
(required to be absolute integrable and bounded). 
The constant $B$ is implicitly determined for a given power budget $S$
by a relation similar to \eqref{eq:capacity:holsinger}.  

Since the time--invariant case represents the commutative setting a fixed signaling scheme (like for
example orthogonal frequency division multiplexing) is 
permitted and the determination of the capacity is essentially reduced to a power allocation problem. Although the
coherent setting (full knowledge at the transmitter) 
is considered so far  only the channel gains have to be given to the transmitter in this case.

However, doubly--dispersive channels represent the non--commutative generalization and do not admit a joint 
diagonalization such that there still remains the problem of proper signal design. Here, the 
correlation operator can be characterized for example by the time--varying transfer function, i.e. the symbol 
$\sigma(x,\omega)$
of a so called pseudo-differential operator $L_\sigma$ which depends on the frequency $\omega$ and the time instant $x$.
Obviously, by uncertainty an exact characterization of frequencies at time instants is meaningless and
the symbol can reflect spectral properties only in an averaged sense. Thus, it is important to know
whether the limit in \eqref{eq:capacity:contin} for a real--valued symbol is asymptotically given by the average:
\begin{equation}
   \frac{1}{\alpha}\iint_{\alpha\Omega\times\Reals}r(B\cdot\sigma(x,\omega))dxd\omega
   \label{eq:capacity:ltv}
\end{equation}
for $\alpha\rightarrow\infty$ and $r(x)=\log(x)\cdot\chi_{[1,\infty)}(x)$. Then, \eqref{eq:capacity:ltv}
with a similar integral with the function $(x-1)/x\cdot\chi_{[1,\infty)}(x)$ represents the water--filling 
principle in time and frequency. Obviously, this strategy is used already in practice when optimizing 
rate functions in some long--term meaning. But, in fast--fading scenarios for example
it not clear whether this procedure on a short time scale is indeed related to  \eqref{eq:capacity:contin}.

Averages closely related to the one in \eqref{eq:capacity:ltv} have been studied for a long time
in the context of asymptotic symbol calculus of pseudo-differential operators and
semi--classical analysis in quantum physics \cite{Widom1982,Hormander1979,Sobolev2010}. Unfortunately, the results
therein are not directly applicable in the information and communication theoretic setting because here
1.) the symbols of the restricted operators are (in general) discontinuous and usually not decaying in 
time 2.) the functions $r$ to be considered are neither analytic nor have the required smoothness
3.) the path of approaching the limit has to be explicitly in terms of an increasing sequence 
of interval restrictions (infinite--dimensional subspaces) in order to establish its operational 
meaning.
For operators with semigroup properties as for example the ''heat channel'' \cite{Hammerich2009}
it is possible to approach the limit via projections onto the (finite--dimensional)
span of an increasing sequence of basis functions (Hermite functions in this case) as established in \cite{Zelditch1989} for 
Schr\"odinger operators. However, in the problem considered here this approach does not guarantees
the existence of signaling schemes of finite length $\alpha$ to practically achieve the limit
and a semigroup property of this particular type is not present.

The idea of approximate eigenfunctions of so called underspread channels \cite{kozek:thesis,jung:wcnc08}
has been used to obtain information--theoretical statements for
the non--coherent setting \cite{Durisi2010}. Signal design has then to be considered 
with respect to statistical properties \cite{jung:wssuspulseshaping}.
The method presented in this paper suggests that in the 
coherent setting the approximation in terms of trace norms is relevant.

\subsection{Main Results}
We establish a procedure for estimating 
the deviation of formula \eqref{eq:capacity:ltv} from the desired quantity \eqref{eq:capacity:contin}.
It will be shown that both terms asymptotically agree for $\alpha\rightarrow\infty$
if the difference of symbol products $L_{\sigma\tau}$ and operator composition
$L_\sigma L_\tau$ can be controlled in trace norm on $\alpha\Omega$ with a sub-linear scaling in $\alpha$.
We will further discuss the information--theoretical impacts: As an example we will study in more
detail symbols $\sigma(x,\omega)$ which are $\Omega$--periodic in $x$. We will show that under certain
integrability and smoothness assumptions on the symbol the limit in \eqref{eq:capacity:contin} is indeed
given as:
\begin{equation}
   I(S)=\iint_{\Omega\times\Reals}r(B\cdot\sigma(x,\omega))dxd\omega
   \label{eq:capacity:ltv:periodic}
\end{equation}
whenever the (inverse) Fourier transform of $\sigma(x,\omega)$ in $\omega$ (the impulse response of 
$L_\sigma$) is supported in a fixed interval.

The paper is organized as follows: In Section \ref{sec:sysmodel} we introduce the channel model
and establish the problem as a Szeg\"o statement on the asymptotic symbol calculus for
pseudo--differential operators. The asymptotic behavior is investigated in Section \ref{sec:asympt:trace}
as a series of four sub--problems: an increasing family of interval sections, the asymptotic symbol calculus, 
an approximation method and finally a result on ''products'' of symbols. Following this line of four
arguments we are able to establish \eqref{eq:capacity:ltv:periodic}.

\section{System Model and Problem Statement}
\label{sec:sysmodel}
We use $L_p(\Omega)$ for usual Lebesgue spaces ($1\leq p\leq\infty$) of complex--valued functions 
on $\Omega\subseteq\Reals^n$ and abbreviate $L_p=L_p(\Reals^n)$ with corresponding norms 
$\lVert\cdot\rVert_{L_p}$. For $p=2$ the Hilbert space has inner product
$\langle u,v\rangle:=\int \bar{u}v$.
Classes of smooth functions up to order $k$ are denoted with $C^k$ and $\fourier{f}=\Fourier f$ 
is the Fourier transform of $f$. Partial derivatives of a function $\sigma(x,\omega)$ are written
as $\sigma_x$ and $\sigma_\omega$, respectively. $\schattenclass_2$ and $\schattenclass_1$ are
Hilbert--Schmidt and trace class operators with square--summable and absolute summable 
singular values and the symbol $\trace(X)$ denotes the trace of
an operator $X$ (more details will be given later on) on $L_2$.
\subsection{System Model}

We consider the common model of transmitting a finite energy signal $s$ with support in an 
interval $\alpha\Omega$ of length $\alpha$ through a channel represented by
a fixed linear operator $H$ and additive distortion $n_k$, i.e. 
quantities measured at the receiver within some interval are expressed as noisy 
correlation responses:
\begin{equation}
   \langle r_k,Hs\rangle+n_k
   \label{eq:tc:channeluse}
\end{equation}
where $\{\langle r_k,\cdot\rangle\}$ are suitable normalized linear functionals implemented at the receiver. 
We assume Gaussian noise with $E(\bar{n}_kn_l)=\langle r_k,r_l\rangle$.

Let us denote with $(Pu)(x)=\chi(x/\alpha)u(x)$ the restriction of a 
function $u$ onto the interval $\alpha\Omega$.
Note that in what follows: \emph{$P$ always depends on $\alpha$}.
We will make in the following the assumption that the restriction $HP$ of the channel operator $H$ to 
input signals of length $\alpha$ with finite energy
is compact, i.e. the restriction $PL_\sigma P$ of the correlation operator $L_\sigma:=H^*H$ is compact as well
($H^*$ denotes the adjoint operator on $L_2$).
This excludes certain channel operators - like the identity -
which are usually referred to as ''dimension-unlimited'', i.e. the wideband cases.
Assume that the kernel $k(x,y)$ of $L_\sigma$ fulfils for all $x\in\Reals$:
\begin{equation}
   |k(x,x-z)|^2\leq\psi(z)
\end{equation}
for some $\sqrt{\psi}\in L_1\cap L_2$\footnote{$\sup_{x\in\Reals} k(x,x-\cdot)\in L_1\cap L_2$}. 
Then its  (Kohn--Nirenberg) symbol or time-varying transfer function
is given by Fourier transformation:
\begin{equation}
   \begin{split}
      \sigma(x,\omega)
      &=\int e^{i2\pi\omega (x-y)} k(x,x-y)dy
   \end{split}
\end{equation}
Throughout the paper we assume that $\sigma$ is real--valued (this can be circumvented 
when passing to the Weyl symbol since $L_\sigma$ is positive--definite). 
It follows that $\lVert \sigma(x,\cdot)\rVert_{L_2}^2\leq \lVert\psi\rVert_{L_1}$ 
uniformly in $x$ and that $L_\sigma$ is bounded on $L_2$:
\begin{equation}
   \begin{split}
      |\langle u, L_\sigma v\rangle|
      &=|\langle u\otimes\bar{v},k\rangle|
      \leq\langle |u\otimes v|,\sqrt{\psi}\rangle\\
      &=\langle |u|,\sqrt{\psi}\ast |v|\rangle
      \leq\lVert\sqrt{\psi}\rVert_{L_1}\lVert u\rVert_{L_2}\lVert v\rVert_{L_2}
   \end{split}
   \label{eq:szego:boundedOp}
\end{equation}
From now on we use $\lVert\cdot\rVert_{\opnorm}:=\lVert\cdot\rVert_{L_2\rightarrow L_2}$ 
to denote the operator norm on $L_2$.
A compact operator $HP$ can be written via the Schmidt representation (singular value decomposition) as a
 limit of a sum of
rank--one operators $HP=\sum_k s_k\langle u_k,\cdot\rangle v_k$ with singular values 
$s_k=\sqrt{\lambda_k(PL_\sigma P)}$ and orthonormal bases
$\{u_k\}$ and $\{v_k\}$ -- \emph{all depending on $\alpha$}. 
For the coherent setting we assume that finite subsets of these 
bases are known and implementable at the transmitter and the receiver, respectively. 
Obviously, this is an idealized and seriously strong assumption which can certainly 
not be fulfilled without error in practise. The investigations in \cite{kozek:identification:bandlimited} suggest
that underspreadness of $H$ is necessary prerequisite for reliable error control.
When representing the signal $s$ as a finite linear combination of $\{u_k\}$ 
a single use of the time--continuous channel $H$ over the time interval  
$\alpha\Omega$ with power budget $S$ is decomposed into a single use of a finite set of time--discrete parallel 
Gaussian channels jointly constrained to $\alpha S$.

We will consider in the following independent uses of the channel in \eqref{eq:tc:channeluse} as our 
preliminary\footnote{We discuss consecutive uses of the same time--continuous channel below.}
model and restrict to $r_k=v_k$, i.e. $E(\bar{n}_kn_l)=\delta_{kl}$. Then, the capacity and the power budget 
of the equivalent memoryless Gaussian channel are related through the water--filling level $B$ 
as (see for example \cite{gallager:inftheo}):
\begin{equation}
   \begin{split}
      \frac{1}{\alpha}\sum_{B\lambda_k\geq 1}\log(B\lambda_k)
      =\frac{1}{\alpha}\traceP r(B\, PL_\sigma P)\\
      \frac{B}{\alpha}\sum_{B\lambda_k\geq 1}\frac{B\lambda_k-1}{B\lambda_k}
      =\frac{B}{\alpha}\traceP p(B\,  PL_\sigma P)\\
   \end{split}
   \label{eq:capacity:restricted}
\end{equation}
with $r(x)=\log(x)\cdot\chi_{[1,\infty)}(x)$ and $p(x)=\frac{x-1}{x}\cdot\chi_{[1,\infty)}(x)$.
The symbol $\traceP Y:=\trace(PYP)$ denotes the trace of the operator $Y$ on the range of $P$ and 
the operators $r(PXP)$ and $p(PXP)$ for $X$ being self--adjoint
are meant by the spectral mapping theorem.

If the time--varying impulse response of $L_\sigma$ (or $H$)
has finite delay ($k(x,x-z)$ is zero for $z$ outside a fixed interval) and is periodic in
the time instants $x$ (the symbol $\sigma(x,\omega)$ is periodic in $x$) multiple channel 
uses in the preliminary model can be taken as consecutive uses
of the same time--continuous channel. Inserting guard periods of appropriate fixed 
size (independent of $\alpha$) will not affect the asymptotic behavior for $\alpha\rightarrow\infty$.
Thus, any further results will then
indeed refer to the information (and coding) capacity. The assumptions on finite delay 
might be relaxed using direct methods like in \cite{cordaro:intersymbol:gaussian} or 
\cite{wyner:intersymbol:gaussian} whereby extensions to almost--periodic channels seems
to lie at the heart of information theory.

\subsection{Problem Statement}
The interval restriction $P$ has the symbol $\chi(x/\alpha)$. The symbol 
of operator products is given as the twisted multiplication of the symbol of the factors. Under the trace
this is reduced to ordinary multiplication (see for example \cite{Estrada1989} in the case of Weyl correspondence).
Thus, the term in \eqref{eq:capacity:ltv} can be written as the following trace:
\begin{equation}
   \frac{1}{\alpha}\traceP L_{f(\sigma)}=\frac{1}{\alpha}\int_{\alpha\Omega\times\Reals} f(\sigma(x,\omega))dxd\omega
   \label{eq:capacity::ltvapprox}
\end{equation}
when taking $f(x)=r(Bx)$. Comparing \eqref{eq:capacity:restricted} with \eqref{eq:capacity::ltvapprox} means
to estimate the asymptotic behavior of:
\begin{equation}
   \begin{split}
      \frac{1}{\alpha}\traceP(f(P L_\sigma P)-L_{f(\sigma)})
   \end{split}
   \label{eq:capacity:error}
\end{equation}
for $\alpha\rightarrow\infty$ (we abbreviate $f(\sigma):=f\circ\sigma$).
As seen from $r$ and $p$ in \eqref{eq:capacity:restricted} the functions 
$f$ of interest are continuous but not differentiable at $x=1$. 

\section{Asymptotic Trace Formulas}
\label{sec:asympt:trace}
The procedure for estimating the difference in \eqref{eq:capacity:error} essentially consists
in the following arguments:
A functional calculus will be used to represent the function $f$ in the operator context.
For $L_{f(\sigma)}$ this can be done independently of $\alpha$ but 
for $f(P L_\sigma P)$ such an approach is much more complicated because of the 
remaining projections $P$. Hence, the first
step is to estimate its deviation to $f(L_\sigma)$ by inserting the zero term 
$\traceP(f(L_\sigma)-f(L_\sigma))/\alpha$ into \eqref{eq:capacity:error}:
\begin{equation}
   \begin{split}
      \frac{1}{\alpha}(\,\overbrace{\traceP [f(PL_\sigma P)-f(L_\sigma)]}^{\text{stability}}+
      \overbrace{\traceP [f(L_\sigma)-L_{f(\sigma)}]}^{\text{symbol calculus}}\,)
   \end{split}
   \label{eq:szego:splitup}
\end{equation}
and use $|\trace(a+b)|\leq |\trace a |+|\trace b |$ to estimate both terms separately. The
first contribution refers to the stability of interval sections (in Section \ref{sec:szego:stability}).
For second term a Fourier--based functional calculus reduces the problem
to the characterization of the approximate product rule for symbols (in Section \ref{sec:szego:asympt:symbol})
which can then be estimated independently of the particular function $f$ (in
Section \ref{sec:szego:approx:product}).
Unfortunately, the last steps require certain smoothness of $f$. Therefore we will approach the 
limit via smooth approximations $f_\epsilon$ as discussed in Section \ref{sec:szego:smooth:approx}.

\subsection{Stability of Interval Sections}
\label{sec:szego:stability}
The following stability result was inspired by the analysis on the Widom conjecture in \cite{Gioev2001}.
Let $\spec(L_\sigma)$ denote
the spectrum of $L_\sigma$. Then the interval $ I:=\bigcup_{t\in[0,1]} t\cdot \spec(L_\sigma)$
contains the spectra of the family $P L_\sigma P$ for each $\alpha$.
\begin{mytheorem}
   Let $L_\sigma$ be an operator with a kernel which fulfils $|k(x,x-z)|^2\leq \psi(z)$ with 
   $\psi\in L_1$. If $\lVert \psi(1-\chi_{[-s,s]})\rVert_{L_1}\leq c/s$
   then:
   \begin{equation}
      \frac{1}{\alpha}|\traceP (f(P L_\sigma P)-f(L_\sigma))|\leq 
      \lVert f''\rVert_{L_\infty(I)}\frac{\log(\alpha)}{\alpha}
   \end{equation}
   for $f\in W^2_\infty(I)$.
   \label{thm:szego:laptev}      
\end{mytheorem}
$W^2_\infty(I)$ denotes the Sobolev class (details in \cite{Laptev1996}). 
Recall that the functions $f$ to be considered here
are continuous and differentiable a.e. on $I$ (except at point $x=1$).
We will shortly discuss the proof of this theorem since it is only a minor variation of
\cite{Gioev2001}.
\begin{proof}
   Laptev and Safarov \cite{Laptev1996} have obtained from Berezin inequality the following
   estimate. For functions $f\in W^2_\infty(I)$ the operator
   $P[f(L_\sigma ) -  f(PL_\sigma P)]P$ is trace class
   if $PL_\sigma $ and $PL_\sigma (1-P)$ are Hilbert--Schmidt with the trace estimate:
   \begin{equation}
      |\traceP (f(L_\sigma )- f(PL_\sigma P))| \leq\frac{1}{2}\lVert f''\rVert_{L_\infty(I)}
      \lVert PL_\sigma (1-P)\rVert^2_{\schattenclass_2}
   \end{equation}
   Recall that the interval projection $P$ is multiplication with the scaled characteristic function $\chi(x/\alpha)$. Thus, change of variables $x=y'+x'$ and $y=y'-x'$ gives:
   \begin{equation}
      \begin{split}
         \lVert PL_\sigma \rVert_{\schattenclass_2}^2
         &=\int\chi(x/\alpha)|k(x,y)|^2  dxdy\\
         &\hspace*{-2em}\leq2\alpha^2\int\psi(2\alpha x')dx'\int\chi(y'+x')dy'\leq\alpha \lVert\psi\rVert_1
      \end{split}
   \end{equation}
   In the same manner we get:
   \begin{equation}
      \begin{split} 
         \lVert PL_\sigma (1&-P)\rVert^2_{\schattenclass_2}
         =\int \chi(\tfrac{x}{\alpha})(1-\chi(\tfrac{y}{\alpha}))|k(x,y)|^2dxdy\\
         &\leq \alpha^2\int \chi(x)(1-\chi(y))\psi(\alpha (x-y))dxdy\\
         &=\alpha^2\int\psi(2\alpha x)\cdot\omega(2x) dx 
      \end{split}
   \end{equation}
   with $\omega(x):=4|x|\leq 2$ for $|x|\leq 1/2$ and $\omega(x):=2$ outside this interval.
   With $u=2\alpha x$ and $\phi(u)=\psi(u)+\psi(-u)$ we split and estimate the integral as follows:
   \begin{equation}
      \begin{split} 
         \lVert PL_\sigma (1&-P)\rVert^2_{\schattenclass_2}    
         =\frac{\alpha}{2}\int_0^\infty \phi(u)\omega(\tfrac{u}{\alpha})du\\
         &\leq\frac{\alpha}{2}\left(\tfrac{8}{\alpha}\int_0^2
           +\int_2^{2\alpha}\tfrac{4u}{\alpha}+2\int_{2\alpha}^\infty \right)\phi(u)du
      \end{split}
   \end{equation}  
   and with the assumptions of the theorem it follows:\vspace*{-.5em}
   \begin{equation}
      \begin{split} 
         \lVert PL_\sigma (1-P)\rVert^2_{\schattenclass_2}    
         &=4\lVert \psi\rVert_1
         +2\int_2^{2\alpha}\phi(u)udu
         +\frac{c}{2}
      \end{split}
      \vspace*{-1em}
   \end{equation}
   Finally we use $\phi(u)=-\frac{d}{du}\int_{u}^{\infty}\phi(s)ds$ and integrate by parts to obtain
   $\int_2^{2\alpha}\phi(u)udu= c (1+\log\alpha)$.
\end{proof}

\subsection{Asymptotic Symbol Calculus}
\label{sec:szego:asympt:symbol}
Here we shall use Fourier techniques to estimate 
the right term in \eqref{eq:szego:splitup}.
We abbreviate in the following $e(x)=\exp(i2\pi x)$. 
\begin{mylemma}
   Let $f$ be a $L_1$-function with $\fourier{f}(\omega)=\Order(\omega^{-4-\delta})$
   for some $\delta>0$.
   For $L_\sigma$ being bounded and self--adjoint on $L_2$ with 
   real--valued symbol $\sigma\in C^{3}$
   it follows that:
   \begin{equation}
      \begin{split}
         \frac{1}{\alpha}|\traceP (f(L_\sigma)-L_{f(\sigma)})|   \leq
         \int dw|\fourier{f}(\omega)|\int_0^\omega Q_\alpha(s)\frac{ds}{\alpha}
      \end{split}
      \label{eq:szego:lemma:asympt:symbol}
   \end{equation}
   with $Q_\alpha(s):=\lVert\left(L_\sigma L_{e(s\sigma )}-L_{\sigma e(s\sigma )}\right)P\rVert_{\schattenclass_1}$.
   \label{lemma:szego:asympt:symbol}
\end{mylemma}
The lemma shows that whenever the rhs in \eqref{eq:szego:lemma:asympt:symbol} is finite
the asymptotics for $\alpha\rightarrow\infty$
is determined only by $Q_\alpha/\alpha$. The function $Q_\alpha$ essentially compares 
the twisted product of $\sigma$ and $e(s\sigma)$ with the ordinary product $\sigma\cdot e(s\sigma )$
in trace norm reduced to intervals of length $\alpha$.
\begin{proof}
   Consider the following operator--valued Bochner integral:
   \begin{equation}
      f(L_\sigma)=\int e(\omega L_\sigma) \fourier{f}(\omega)d\omega
      \label{eq:szego:thm:gL}
   \end{equation}
   where the operator $e(\omega L_\sigma)$
   is defined as the usual power series converging in norm since $L_\sigma$ is bounded. In
   particular $e(\omega L_\sigma)$ is unitary on $L_2$ ($L_\sigma$ is self--adjoint) 
   and depends continuously on $\omega$. 
   Since 
   $\lVert f(L_\sigma)\rVert_{\opnorm}\leq\lVert \fourier{f}\rVert_1$
   convergence in operator norm is guaranteed and the construction agrees with the spectral
   mapping theorem (see \cite{Taylor1968}).
   The value of the symbol $f\circ\sigma$ at each point can be expressed in terms of $\fourier{f}$. This suggests
   the formula:
   \begin{equation}
      L_{f(\sigma)}=\int L_{e(\omega\sigma)}\fourier{f}(\omega)d\omega
      \label{eq:szego:thm:Lg}
   \end{equation}
   From Calderon Vaillancourt Theorem \cite[Ch.5]{folland:harmonics:phasespace} 
   we have:
   \begin{equation}
      \lVert L_{e(s\sigma)}\rVert_\opnorm\leq \lVert e(s\sigma)\rVert_{C^3}
      :=\sum_{a+b\leq 3} |2\pi s|^{a+b}\lVert\partial_x^a\partial_\omega^b\sigma\rVert_{L_\infty}
      \label{eq:szego:thm:LeNorm}
   \end{equation}
   Thus, for $\fourier{f}(\omega)=\Order(\omega^{-4-\delta})$ and $\delta>0$ the integral \eqref{eq:szego:thm:Lg}
   converge in the sense of Bochner. 
   From the considerations above we get therefore:
   \begin{equation}
      \begin{split}
         |\traceP (L_{f(\sigma)}-f(L_\sigma))|
         &\leq\int |\fourier{f}(\omega)|\cdot
         |\traceP u(\omega)|d\omega
      \end{split}
   \end{equation}
   with $u(\omega)=L_\sigma e(\omega L_\sigma)-L_{\sigma e(\omega\sigma)}$.
   As suggested in \cite{Sobolev2010} the operator $u(\omega)$ fulfils the following 
   identity\footnote{in case of operators: 
   $\partial_\omega e(\omega L_\sigma)=i2\pi L_\sigma  e(\omega L_\sigma)$ \cite[Lemma 5.1]{Gohberg1990}.}:
   \begin{equation}
      \begin{split}
         u'(\omega)
         &=i2\pi\left(L_\sigma  u(\omega)+L_\sigma L_{e(\omega\sigma)}-L_{\sigma e(\omega\sigma)}\right)\\
      \end{split}
   \end{equation}
   i.e. an inhomogenous Cauchy problem with initial condition  $u(0)=0$.
   By Duhamel's principle (see for example
   \cite[p.50]{Cazenave1998} for the Banach--space valued case):
   \begin{equation}
      \begin{split}
         \hspace*{-1em}u(\omega)=\frac{2\pi}{i}\int_0^\omega e((\omega-s)L_\sigma )
         \left(L_\sigma L_{e(s\sigma )}-L_{\sigma e(s\sigma )}\right)ds
      \end{split}
   \end{equation}
   giving the estimate:
   \begin{equation}
      \begin{split}
         |\traceP u(\omega)|\leq\int_0^\omega
         \lVert\left(L_\sigma L_{e(s\sigma )}-L_{\sigma e(s\sigma )}\right)P\rVert_{\schattenclass_1}ds
      \end{split}
   \end{equation}
   since $\lVert Pe((t-s)L_\sigma )\rVert_\opnorm\leq 1$. 
\end{proof}
The smoothness assumptions in the theorem can be weakened to $\sigma\in C^{2+\delta}$ and 
$\fourier{f}(\omega)=\Order(\omega^{-3-\delta})$ when using H\"older-Zygmund spaces. 
We expect that these conditions can be further reduced when
using in \eqref{eq:szego:thm:Lg} some weaker convergence in $\traceP$ 
instead of requiring a Bochner integral. The proof of the theorem can also be based
on the Paley--Wiener theorem, i.e. 
$f\rightarrow f(L_\sigma)$ and $f\rightarrow L_{f(\sigma)}$ are operator--valued
distributions of compact support with order at most $3$  and have therefore 
$C^3$ as natural domain.

\subsection{An Approximation Procedure}
\label{sec:szego:smooth:approx}
Since $L_\sigma$ is bounded (see \eqref{eq:szego:boundedOp}) the functions $f$ 
will be evaluated only on a finite interval contained in $I$. We consider 
functions $f$  of the form $f(x)=h(x)\cdot\chi_{[1,\infty)}(x)$
with a critical point at $x=1$.
By smooth extension outside the interval its Fourier transforms $\fourier{f}(\omega)$ decay 
only as $\Order(\omega^{-2})$,  
see here for example \cite[Theorem 2.4]{Trefethen1996}, i.e. $f\in L_1\cap\Fourier L_1$.
Unfortunately, this is not sufficient for Lemma \ref{lemma:szego:asympt:symbol}.
Therefore, we replace the 
Heaviside function $\chi_{[1,\infty)}$
in $f$ by a series of smooth approximations $\phi_\epsilon$ as done for example in \cite{Hormander1979}.
Let be $\phi\in C^\infty$ with $\phi(t)=0$ for $t\leq 0$ and $\phi(t)=1$ for
$t\geq 1$. Define $\phi_\epsilon(x)=\phi(\frac{x-1}{\epsilon})$ and consider  
$f_\epsilon=h\phi_\epsilon\in C^\infty_c$ (a smooth function of compact support,
achieved again by smooth extension outside the interval $I$)
instead of $f$:
\begin{equation}
   \begin{split}
      |\fourier{f}_{\epsilon}(\omega)|
      \leq\frac{c'_n|I|}{|2\pi\omega|^{n}}\epsilon^{-n}
   \end{split}
   \label{eq:szego:smoothapprox:fourier:decay}
\end{equation}
In essence: polynomial grow of $Q_\alpha(s)$ in $s$ can always be compensated 
by taking $n$ large enough such that at the rhs in \eqref{eq:szego:lemma:asympt:symbol} remains
a finite quantity $R_\alpha(\epsilon)$. If for example $R_\alpha(\epsilon)=\Order(\alpha^{-\gamma})$, 
we choose $\epsilon=\alpha^{-\delta}$ with $\delta<\gamma/n$. Then 
$R_\alpha(\epsilon)\rightarrow 0$ and $\epsilon\rightarrow 0$ for $\alpha\rightarrow\infty$
which is obviously sufficient for the limit.

\subsection{Approximate Symbol Products}
\label{sec:szego:approx:product}
Let us abbreviate $\tau=e(s\sigma )$. Then the operator in the term $Q_\alpha(s)/\alpha$
of Lemma \ref{lemma:szego:asympt:symbol} is the deviation between operator and symbol product
$L_\sigma L_{\sigma \tau}-L_{\sigma \tau}$. 
As in \cite{Widom1982} we insert $L_\sigma L_{\bar{\tau}}^*-L_\sigma L_{\bar{\tau}}^*$ and apply
triangle inequality to obtain:
\begin{equation}
   Q_\alpha(s)\leq
   \lVert L_\sigma\rVert_{\opnorm}\lVert TP\rVert_{\schattenclass_1}
   +\lVert T'P\rVert_{\schattenclass_1}
\end{equation}
where $T=L_{\bar{\tau}}^*-L_{\tau}$ and $T'=L_\sigma L_{\bar{\tau}}^*-L_{\sigma \tau}$ having 
kernels $t(x,y)$ and $t'(x,y)$:
\begin{equation}
   \begin{split}
      t(x,y)&=\int e^{i2\pi(x-y)\omega}(\tau(x,\omega)-\tau(y,\omega))d\omega\\
      t'(x,y)&=\int e^{i2\pi(x-y)\omega}\sigma(x,\omega)(\tau(x,\omega)-\tau(y,\omega))d\omega
   \end{split}
   \label{eq:szego:tkernels}
\end{equation}
Polynomial orders in $s$ which will occur in the following will be compensated by the approximation method
in Section \ref{sec:szego:smooth:approx}. The role of $\tau$ and $\sigma$ can also be interchanged 
since according \eqref{eq:szego:thm:LeNorm} $L_\tau$ is bounded polynomially in $s$.

We will discuss in the following 
under which conditions $\lVert TP\rVert_{\schattenclass_1}$ is finite and what will be scaling in $\alpha$. 
The argumentation for $\lVert T'P\rVert_{\schattenclass_1}$ will be analogous.
From integration by parts (since $\tau(x,\omega)-\tau(y,\omega)\rightarrow 0$ for $|\omega|\rightarrow\infty$)
we have:
\begin{equation}
   \begin{split}
      t(x,y)&=\frac{h(x,x-y)-h(y,x-y)}{i2\pi(x-y)}
   \end{split}
\end{equation}
where $h(x,z)=\int e^{i2\pi\omega z}\tau_\omega(x,\omega)d\omega$ and 
$|\tau_\omega|=|2\pi s\sigma_\omega|$. It is already assumed that
$\sigma(x,\cdot)\in L_2$ uniformly in $x$. If $\sigma_\omega(x,\cdot)$ is of bounded variation
(or even continuous) we deduce with the mean value theorem
that $|t(x,y)|^2\leq c/(1+|x-y|^2)$. This in turns implies that
$\lVert TP\rVert_{\schattenclass_2}=\Order(\sqrt{\alpha})$ but it will not be sufficient
for $\lVert TP\rVert_{\schattenclass_1}$. 

It is known that
general trace class estimates can not be achieved in this way 
and further smoothness assumptions are necessary.
The problem is related to the absolute summability of orthogonal 
series (in particular Fourier series as shown later on) which is evident from:
\begin{equation}
   \begin{split}
      \lVert TP\rVert_{\schattenclass_1}
      &\leq \sum_n\lVert T\phi_n\rVert_{L_2}
   \end{split}
\end{equation}
where $\phi_n$ is an ONB for the range of $P$ ($\text{supp}\,\phi_n\subseteq\alpha\Omega$). 
However, for finite--rank $TP$ it follows here already from
$\lVert T\phi_n\rVert_{L_2}\leq \lVert TP\rVert_{\schattenclass_2}=\Order(\sqrt{\alpha})$
that the approximation method
in Section \ref{sec:szego:smooth:approx} can be applied giving the correct statement in
\eqref{eq:capacity:error}. 

Let be $c_n$ a (positive) sequence $1/c_n\rightarrow0$ as $n\rightarrow\infty$ with
$K=\sum_n c_n^{-2\lambda}$ being finite for $\lambda>1/2$. 
H\"older inequality implies:
\begin{equation}
   \begin{split}
      \lVert TP\rVert^2_{\schattenclass_1}
      &\leq K \sum_n \lVert c_n^\lambda T\phi_n\rVert^2_{L_2}\\
      &=K\int_\Reals \sum_n|\langle \bar{t}(z+\cdot,\cdot),c^\lambda_n\phi_n\rangle|^2 dz\\
   \end{split}
   \label{eq:szego:ortho:sum}
\end{equation}
\emph{The periodic case:}
Let us assume that $t(z+y,y)$ is periodic in $y$ (same for $t'$) with period $1$ 
(for simplicity) which is given 
if the symbol $\sigma(x,\omega)$ is $1$--periodic in $x$. We use
the Fourier basis  $\phi_n(y)=\exp(i2\pi ny/\alpha)/\sqrt{\alpha}$  and consider 
$\alpha\in\Naturals$ and $\Omega=[0,1]$, i.e. $t$ is given by the series:
\begin{equation}
   t(z+y,y)=\sqrt{\alpha}\sum_m \fourier{t}_m(z)\phi_{\alpha m}(y)
\end{equation}
The sum in \eqref{eq:szego:ortho:sum} reduces to the indexes $\alpha n$.
We take exemplary $\lambda=1$, $c_{\alpha n}=2\pi n$  and $c_{\alpha n+k}=2\pi\sqrt{\alpha-1}n$ for $k=1\dots\alpha-1$
such that $K$ is independent of $\alpha$.
Then \eqref{eq:szego:ortho:sum} is:
\begin{equation}
   \begin{split}
      \lVert TP\rVert_{\schattenclass_1}\leq  
      \sqrt{\alpha}\left(K\int \sum_m |2\pi m \fourier{t}_m(z)|^2 dz\right)^{1/2}
   \end{split}
   \label{eq:szego:ortho:sum:fourier}
\end{equation}
which results in the condition $t_y\in L_2(\Reals\times \Omega)$
(to be precise, the derivative in the $L_2$--mean - see \cite{Stinespring1958}). 
From the definition
of $t$ in \eqref{eq:szego:tkernels} it follows that:
\begin{equation}
   \begin{split}
      t_y(z+y,y)=\int e^{i2\pi z\omega}\left(\tau_x(y+z,\omega)-\tau_x(y,\omega)\right)d\omega
   \end{split}
   \label{eq:szego:tykernel}
\end{equation}
where $|\tau_x|=2\pi s|\sigma_x|$.
Thus, the latter condition is fulfilled if $\sigma_x(x,\cdot)\in L_1$ and the derivative in $\omega$ has
bounded variation (or is continuous). Thus, in this case the method in Section \ref{sec:szego:smooth:approx} 
can be applied which proves formula \eqref{eq:capacity:ltv:periodic}.

\section{Conclusion}
A new approach to the capacity of time--continuous doubly--dispersive Gaussian channels
with periodic symbol has been established by proving an asymptotic Szeg\"o result  
for certain pseudo--differential operators.

\section*{Acknowledgment}
This work is supported by Deutsche Forschungsgemeinschaft (DFG) grant JU 2795/1-1.



\bibliographystyle{IEEEtrannourl}
\bibliography{library}

\end{document}